\documentclass[prd,onecolumn]{revtex4}
\usepackage{dcolumn}
\usepackage{multirow}
\usepackage{graphicx}
\usepackage{amssymb}
\usepackage{bm}
\usepackage{hyperref}%for .pdf
\usepackage{epstopdf}%for .pdf
\usepackage{color}
\usepackage{mathrsfs}
\usepackage{amsmath,amssymb,amsthm}
\begin{document}
\title{A pressure parametric dark energy model}
\author{Jun-Chao Wang$^{1}$}
\email{dakaijun@dakaijun.cn}
\author{Xin-He Meng$^{1}$}
\email{xhm@nankai.edu.cn}
\affiliation{
$^1${Department of Physics, Nankai University, Tianjin 300071, China}\\}
\begin{abstract}
In this paper, we propose a new pressure parametric model of the total cosmos energy components in a spatially flat Friedmann-Robertson-Walker (FRW) universe and then reconstruct the model into quintessence and phantom scenarios, respectively. By constraining with the datasets of the type Ia supernova (SNe Ia), the baryon acoustic oscillation (BAO) and the observational Hubble parameter data(OHD), we find that $\Omega_{m0}=0.270^{+0.039}_{-0.034}$ at the 1$\sigma$ level and our universe slightly biases towards quintessence behavior. Then we use two diagnostics including $Om(a)$ diagnostic and statefinder to discriminate our model from the cosmology constant cold dark matter ($\Lambda$CDM) model. From $Om(a)$ diagnostic, we find that our model has a relatively large deviation from the $\Lambda$CDM model at high redshifts and gradually approaches the $\Lambda$CDM model at low redshifts and in the future evolution, but they can be easily differentiated from each other at the 1$\sigma$ level all along. By the statefinder, we find that both of quintessence case and phantom case can be well distinguished from the $\Lambda$CDM model and will gradually deviate from each other. Finally, we discuss the fate of universe evolution (named the rip analysis) for the phantom case of our model and find that the universe will run into a little rip stage.
\end{abstract}
\maketitle
%Keywords: xxx, xxx.
\section{Introduction}
From the conventional Einstein field equation dominated by matter without negative pressure ($G_{\mu\nu}=8\pi GT_{\mu\nu}$) and Hubble law, it can be concluded that the universe is in a decelerating expansion period, but since the reported result of the accelerated expansion of the universe from the supernova data observed in 1998 and 1999\cite{riess1998observational,perlmutter1999constraining}, there have been continuous data to prove that the current universe is in the phase of accelerated expansion. In order to accommodate this phenomenon, one way is to modify the left side of the traditional Einstein field equation (modify the gravity). Another way is to add a negative pressure matter component named dark energy to the right side of the equation. One of the global fitting well scenarios is the so-called standard cosmology or the $\Lambda$CDM model which includes the simplest dark energy model with the equation of state (EoS) $\omega\equiv\frac{p}{\rho}=-1$ that provides a reasonably good account of the properties of the currently observed cosmos such as accelerating expansion of the universe, the large-scale structure and cosmic microwave background (CMB) radiation. However, there has a major outstanding problem (named the fine-tuning problem) that the observed value of dark energy density is 120 orders of magnitude smaller than the theoretical value in quantum field theory if taking the allowed highest energy cut off scale as the Planck mass\cite{carroll2001cosmological,weinberg1989cosmological}; besides, there is also the so called coincidence problem which asks why dark energy density and physical material density are exactly in the same order of magnitude. To alleviate these problems, some extended models have been raised such as an evolving scalar field with the time variant EoS, for example.

In order to study the characterization of dark energy component, one of the feasible methods is to parameterize some observable physical quantities and then use the observed data to quantify the parameters. The  mainstream is the EoS parametrization, such as Chevalier-Polarski-Linder (CPL)\cite{chevallier2001accelerating,linder2003exploring} parametrization $\omega_{de}(z)=\omega_{0}+\frac{\omega_{a}z}{1+z}$ which behaves as $\omega_{de}\rightarrow\omega_{0}$ for $z \rightarrow 0$ and $\omega_{de}\rightarrow\omega_{0}+\omega_{a}$ for $z \rightarrow \infty$. A few years later a more general form $\omega_{de}(z)=\omega_{0}+\frac{\omega_{a}z}{(1+z)^{p}}$ named Jassal-Bagla-Padmanabhan (JBP)\cite{jassal2005wmap} parametrization has been proposed. In addition, C. Wetterich\cite{wetterich2004phenomenological} has also given a parametric form which goes by $\omega_{de}(z)=\frac{\omega_{0}}{[1+b\ln (1+z)]^{2}}$ and it behaves as $\omega_{de}\rightarrow\omega_{0}$ for $z \rightarrow 0$ and $\omega_{de}\rightarrow 0$ for $z \rightarrow \infty$.

In recent years, some pressure parametric models for the mysterious dark energy or total energy components have been continuously proposed. In 2008, A.A Sen, S. Kumar and A. Nautiyal\cite{sen2008deviation,kumar2013deviation} have put forward a pressure parametric model of dark energy $P_{\Lambda}=-P_{0}+P_{1}\frac{z}{1+z}+...$. Seven years later, Q. Zhang, G. Yang, Q. Zou, X. Meng, K. Shen and D. Wang \cite{zhang2015exploring,yang2016diagnostics} have proposed two dark energy models for the total pressure $P(z)=P_{a}+P_{b}z$ and $P(z)=P_{c}+P_{d}\frac{z}{1+z}$. Then, two years latter, D. Wang, Y. Yan and X. Meng\cite{wang2017new} have raised a pressure-parametrization unified dark fluid model $P(z)=P_{a}+P_{b}(z+\frac{z}{1+z})$.  In the following of this paper we give out a new pressure parametric model of the total energy components as $P(z)=P_{a}+P_{b}\ln (1+z),(z\neq -1)$ in a spatially flat FRW universe and then we discuss its property detailedly.

To investigate the model properties in details, this paper is organized as follows: In Sec. \ref{sec2}, we propose the parametric model by continuously previous studying with the essential formalism and discuss the meanings for the two parameters of the model analytically. Sec. \ref{sec4} is the reconstructions of our model with the quintessence and phantom scalar fields, respectively. In Sec. \ref{sec5}, we constrain our model by using data from SNe Ia, BAO and OHD. In Sec. \ref{sec6}, we discriminate our model from the $\Lambda$CDM model by using $Om(a)$ diagnostic and the statefinder parameters. Sec. \ref{sec7} shows the discussions about the fate of universe evolution named the rip analysis for the phantom case of proposed model. In the last section, Sec. \ref{sec8}, the conclusions and discussions are given.
\section{The parametric model}\label{sec2}
Though two decades have passed a consistent and convincing dark energy theory is yet to come. To understand the puzzling dark energy physics better and by keeping on our exploration, we can properly parameterize its pressure. For example, one can hypothesize a relation between the pressure and the redshift, then integrate out the expression of the density $\rho$ through the conservation equation. Finally from the Fridemann equations $H^{2}=\frac{8\pi G}{3}\sum_{i}\rho_{i}$ and the EoS $\omega=\frac{P}{\rho}$ we are able to get the form of the Hubble parameter $H$ and $\omega$ expression, respectively. By this treatment  so far, a closed system for the evolution of the universe has been established which is described by the Friedmann equations, the conservation or continous equation and the EoS form.

 Assume a relationship between the pressure of all energy components in the universe and the redshift as below,
\begin{equation}
P(z)=P_{a}+P_{b}\ln (1+z),(z\neq -1), \label{1}
\end{equation}
where $P_{a}$ and $P_{b}$ are free parameters. We make this assumption because the form of $\ln(1+z)=z-\frac{z^{2}}{2}+\frac{z^{3}}{3}-...$ for $|z|<1$ and it may be much helpful for providing more opportunities to further other studies related. When $P_{b}=0$, the model is reduced to the well known $\Lambda$CDM model; while when $P_{b}\neq 0$, the total pressure gives more interesting properties. By using the relation of scale factor $a=\frac{1}{1+z}$ and the conservation equation $\dot{\rho }+3\frac{\dot{a}}{a}(P+\rho ) = 0$,  we have 
\begin{equation}
\rho (a)=-(P_{a}+\frac{1}{3}P_{b})+P_{b}\ln a +Ca^{-3}, \label{2}
\end{equation}
where $C$ is an integration constant. We assume that $\rho_{0}$ is the present-day energy density i.e. $\rho (a=1) = \rho_{0}$. Finally the total energy density and pressure can be integrated separately as
\begin{equation}
\rho (a)=\rho_{0}(1-\Omega_{m0}-\alpha \ln a+\Omega_{m0}a^{-3}),\label{3}
\end{equation}
\begin{equation}
P (a)=\rho_{0}(-1+\Omega_{m0}+\frac{1}{3}\alpha+\alpha \ln a).\label{4}
\end{equation}
Here the parameters $(P_{a},P_{b})$ are replaced by new dimensionless parameters $(\alpha,\Omega_{m0})$ where $\alpha\equiv -\frac{P_{b}}{\rho_{0}}$ and $\Omega_{m0} \equiv \frac{1}{\rho_{0}}(\rho_{0}+P_{a}+\frac{1}{3}P_{b})$.

In this model, $\rho(a)$ contains cosmic matter contribution $\Omega_{m0}a^{3}$ and the cosmic dark energy composition $1-\Omega_{m0}-\alpha \ln a$. If we require that the density of each component would be greater than zero, then $\frac{\rho_{de}}{\rho_{0}}=1-\Omega_{m0}+\alpha\ln \frac{1}{a}>0$, so the part of $a>\exp(\frac{1-\Omega_{m0}}{\alpha})>1$ if $\alpha>0$ and  $a<\exp(\frac{1-\Omega_{m0}}{\alpha})<1$ if $\alpha>0$ are out of discussing. Further, if we bend the rules and only require $H^{2}=\frac{8\pi G}{3}\rho_{0}(1-\Omega_{m0}-\alpha \ln a+\Omega_{m0}a^{-3})$ to be greater than zero, then, for $\alpha>0$, $H^{2}$ goes less than zero with a large $a$; For $\alpha<0$, $H^{2}$ increases first and then decreases, and gets the minimum at $a=(-\frac{3}{\alpha}\Omega_{m0})^{\frac{1}{3}}$. Taking an example of $\Omega_{m0}=0.3$, the $\alpha>-4$ guarantees $H^{2}>0$. The EoS of the dark energy and the dimensionless Hubble parameter take the form, respectively
\begin{equation}
\omega_{de}=\frac{P_{de}}{\rho_{de}}=-1+\frac{\frac{1}{3}\alpha}{1-\Omega_{m0}-\alpha \ln a}.\label{5}
\end{equation}
\begin{equation}
E(a)^{2}=1-\Omega_{m0}-\alpha \ln a+ \Omega_{m0}a^{-3}.\label{6}
\end{equation}
To exhibit dark energy better, we derive the density ratio parameter of the dark energy as follows
\begin{equation}
\Omega_{de}=\frac{1-\Omega_{m0}-\alpha \ln a}{1-\Omega_{m0}-\alpha \ln a +\Omega_{m0}a^{-3}}.\label{7}
\end{equation}

Two parameters $\alpha$ and $\Omega_{m0}$ will be constrained by observations in the following section. On the one hand, when $a=1$, $\Omega_{m}=1-\Omega_{de}=\Omega_{m0}.$ So $\Omega_{m0}$ is the present-day dark matter density parameter. On the other hand, when $\alpha=0$, the model reduces to  the flat $\Lambda$CDM model. Further, we can see clearly that in the next section, the $\alpha>0$ for the quintessence case while $\alpha<0$ corresponded to the phantom case.
\section{The reconstructions}\label{sec4}
Unlike the $\Lambda$CDM model, this scenario gets the dynamical dark energy within. The natural way to introduce varying dark energy is to assume a scalar field that changes over time and the corresponding pressure and energy are respective i.e. $P_{de}=P_{scalar}$, $\rho_{de}=\rho_{scalar}.$ In this section, we discuss the quintessence and phantom scalar field separately. Consider the dark energy as a real scalar field $\phi$ with the action of stress energy which can be written as
\begin{equation}
S_{\phi}=-\int \mathrm{d}^{4}x\sqrt{-g}\left[\frac{b}{2}\partial_{\mu}\phi\partial^{\mu}\phi +V(\phi)\right],\label{8}
\end{equation}
where $\frac{b}{2}g^{\mu\nu}\partial_{\mu}\phi\partial_{\nu}\phi$ is the kinetic energy and $V(\phi)$ is the potential energy, $b=1$ or $-1$ corresponding to the quintessence case and phantom case, respectively. And the stress–energy tensor is
\begin{equation}
T^{\mu\nu}(\phi)=\partial^{\mu}\phi\partial^{\nu}\phi-g^{\mu\nu}\left[\frac{b}{2}\partial_{\alpha}\phi\partial^{\alpha}\phi +V(\phi)\right].\label{9}
\end{equation}
If we regard the scalar field as a perfect fluid, the energy density and pressure of the scalar field can be written as
\begin{equation}
\rho_{\phi}=-\frac{b}{2}g^{\mu\nu}\partial_{\mu}\phi\partial_{\nu}\phi+V(\phi),
\label{10}
\end{equation}
\begin{equation}
P_{\phi}=-\frac{b}{2}g^{\mu\nu}\partial_{\mu}\phi\partial_{\nu}\phi-V(\phi).
\label{47}
\end{equation}
Assume $\phi$ is uniform in space and only relies on time i.e. $\phi=\phi(t)$, then Eqs.(\ref{10}) and (\ref{47}) can be simplified to 
\begin{equation} 
\rho_{\phi}=\frac{b\dot{\phi}^{2}}{2}+V(\phi),
\label{11}
\end{equation}
\begin{equation} 
P_{\phi}=\frac{b\dot{\phi}^{2}}{2}-V(\phi),
\label{12}
\end{equation}
where the dot denotes the derivatives w.r.t. the cosmic time.
\subsection{The quintessence case}
Assume the universe consists of quintessence and matter. By comparing Eqs.(\ref{3}) and (\ref{4}) with Eqs.(\ref{11}) and (\ref{12}), we can obtain
\begin{equation} 
\rho_{de}=\frac{1}{2}\dot{\phi}^{2}+V(\phi)=\rho_{0}(1-\Omega_{m0}-\alpha \ln a),
\label{13}
\end{equation}
\begin{equation} 
P_{de}=\frac{1}{2}\dot{\phi}^{2}-V(\phi)=\rho_{0}(\Omega_{m0}-1+\frac{1}{3}\alpha +\alpha \ln a).
\label{14}
\end{equation}
Simplify the above two equations, then we have
\begin{equation} 
\dot{\phi}^{2}=\frac{1}{3}\rho_{0}\alpha,
\label{15}
\end{equation}
\begin{equation} 
V=\rho_{0}(1-\Omega_{m0}-\alpha\ln a-\frac{\alpha}{6}),
\label{16}
\end{equation}
\begin{equation} 
\frac{d\phi}{da}=(aH)^{-1}\dot{\phi}=\pm\frac{M_{pl}}{a}\sqrt{\frac{\alpha}{1-\Omega_{m0}-\alpha\ln a + \Omega_{m0}a^{-3}}},
\label{17}
\end{equation}
where $M_{pl}\equiv(8\pi G)^{-1/2}$ and $H^{2}=\frac{\rho}{3M_{pl}^{2}}$. In Eq.(\ref{17}), '$\pm$' corresponds to two solutions. Only when $\alpha>0$, Eq.(\ref{17}) is meaningful. So $\alpha>0$ corresponds to the quintessence case. And from Eq.(\ref{5}) we know that at this time $\omega_{de}>-1$.

\begin{figure}
\centering
\includegraphics[scale=0.5]{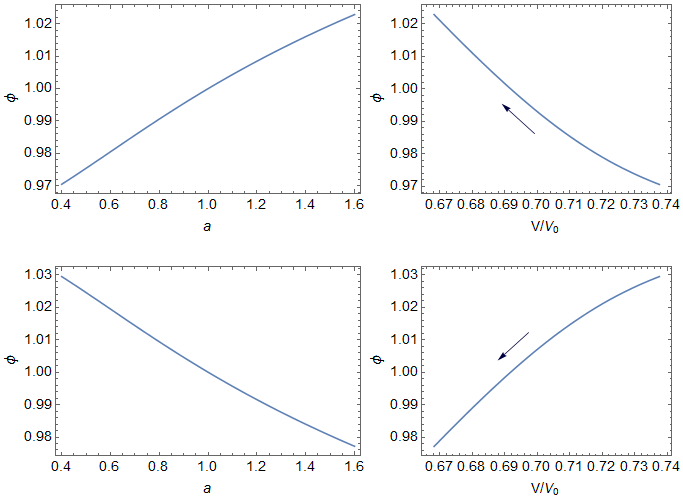}
\caption{The quintessence field $\phi$ versus the scale factor $a$, and the quintessence field $\phi$ versus the potential $V/V_{0}$ (assume $V_{0}=\rho_{0}$). The upper and lower panels correspond to the plus and minus sign in Eq.(\ref{17}), respectively. The arrows indicate the evolutional directions of the potential, and we have used $\Omega_{m0}=0.3$ and $\alpha=0.05$ numerically. }\label{f1}
\end{figure}

From Eqs.(\ref{16}) and (\ref{17}) we can draw the relation between $\phi(a)$ and $V(\phi)$ shown in Fig.\ref{f1}. In Fig.\ref{f1}, from the upper panels we know that $\phi$ increases as $a$ increases while $V$ decreases as $\phi$ increases. The lower panels show that $\phi$ decreases as $a$ increases while $V$ decreases as $\phi$ decreases. So for quintessence case, $V$ decreases as $a$ increases and it implies $\rho_{de}$ will decrease in the future. 
\subsection{The phantom case}
Assume the universe consists of phantom and matter. By comparing Eqs.(\ref{3}) and (\ref{4}) with Eqs.(\ref{11}) and (\ref{12}), we can obtain
\begin{equation} 
\rho_{de}=-\frac{1}{2}\dot{\phi}^{2}+V(\phi)=\rho_{0}(1-\Omega_{m0}-\alpha \ln a),
\label{18}
\end{equation}
\begin{equation} 
P_{de}=-\frac{1}{2}\dot{\phi}^{2}-V(\phi)=\rho_{0}(\Omega_{m0}-1+\frac{1}{3}\alpha +\alpha \ln a).
\label{19}
\end{equation}
Subsequently, by solving the above two equations, one can derive
\begin{equation} 
\dot{\phi}^{2}=-\frac{1}{3}\rho_{0}\alpha,
\label{20}
\end{equation}
\begin{equation} 
V=\rho_{0}(1-\Omega_{m0}-\alpha\ln a-\frac{\alpha}{6}),
\label{21}
\end{equation}
\begin{equation} 
\frac{d\phi}{da}=(aH)^{-1}\dot{\phi}=\pm\frac{M_{pl}}{a}\sqrt{-\frac{\alpha}{1-\Omega_{m0}-\alpha\ln a + \Omega_{m0}a^{-3}}},
\label{22}
\end{equation}
where $M_{pl}\equiv(8\pi G)^{-1/2}$ and $H^{2}=\frac{\rho}{3M_{pl}^{2}}$. In Eq.(\ref{22}), '$\pm$' corresponds to two solutions. Only when $\alpha<0$, Eq.(\ref{22}) is meaningful. So $\alpha<0$ corresponds to the phantom case. From Eq.(\ref{5}) we know at this time $\omega_{de}<-1$.

\begin{figure}
\centering
\includegraphics[scale=0.5]{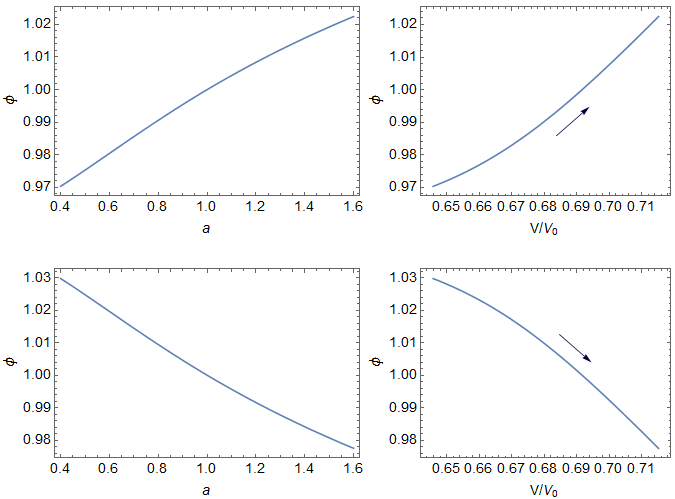}
\caption{The phantom field $\phi$ versus the scale factor $a$, and the phantom field $\phi$ versus the  potential $V/V_{0}$  (Assume $V_{0}=\rho_{0}$). The upper and lower panels correspond to the plus and minus sign in Eq.(\ref{22}), respectively. The arrows indicate the evolutional directions of the potential. We have used $\Omega_{m0}=0.3$ and $\alpha=-0.05$ numerically.}\label{f2}
\end{figure}

From Eqs.(\ref{21}) and (\ref{22}) we can draw the relation between $\phi(a)$ and $V(\phi)$ shown in Fig.\ref{f2}. In Fig.\ref{f2}, from the upper panels we know that $\phi$ increases as $a$ increases while $V$ increases as $\phi$ increases. And the lower panels show that $\phi$ decreases as $a$ increases while $V$ increases as $\phi$ decreases. So for the phantom case, $V$ increases as $a$ increases which implies $\rho_{de}$ will increase in the future and lead $H\rightarrow\infty$ as $a\rightarrow\infty$, and universe will get rip in the end.
\section{The constraints}\label{sec5}
\subsection{Type Ia Supernova}
Measuring the distance by the light curve of a supernova is one of the most accurate ways to measure the distance to the universe. In this paper we use the Union2.1 SNe Ia dataset\cite{suzuki2012hubble}, which contains 580 SNe Ia. First, we minimize the chi-square
\begin{equation} 
\chi^{2}_{SN}=\sum_{i=1}^{580}\frac{\left[\mu_{obs}(z_{i})-\mu(z_{i}) \right]^2}{\sigma_{i}^{2}},
\label{23}
\end{equation}
where $\mu_{obs}(z_{i})$ is the observed distance modulus, $\sigma_{i}$ is the 1$\sigma$ level of the observed distance modulus for each supernova and $\mu(z_{i})$ is the theoretical distance modulus which is defined as 
\begin{equation}
\mu_{z}=5\ln \frac{D_{L}}{H_{0}}+C=5\ln D_{L}-5 \ln H_{0}+C,
\label{24}
\end{equation}
where $H_{0}$ is the Hubble parameter at $z=0$, $C$ is the zero value of the distance modulus and $D_{L}$ is the Hubble-free luminosity distance in a spatially flat FRW universe which can be written as
\begin{equation}
D_{L}=(1+z)\int_{0}^{z}\frac{dz'}{E(z')},
\label{25}
\end{equation}
where $E(z)$ is the dimensionless Hubble parameter. Since the zero value $C$ in Eq.(\ref{24}) of the distance modulus measured in the astronomical observation is arbitrarily selected, $H_{0}$ is also arbitrary. In Eq.(\ref{24}), $H_{0}$ appears in $5\ln H_{0}$. Assume $x=5\ln H_{0}$ for a uniform distribution, $P(x)=1$. Then the likelihood for marginalize $x$ can be written as
\begin{equation}
\exp (-\widetilde{\chi}_{SN}^{2}/2)=\int \exp (-\chi_{SN}^{2}/2)P(x)dx.
\label{26}
\end{equation}
By solving eq.(\ref{26}), we get the marginalized result
\begin{equation}
\widetilde{\chi}_{SN}^{2}=\sum_{i=1}^{580}\frac{\mu_{i}^{'2}}{\sigma_{i}^{2}}-\frac{(\sum _{i=1}^{580}\mu'_{i}/\sigma_{i}^{2})^{2}}{\sum_{i=1}^{580}1/\sigma_{i}^{2}},
\label{27}
\end{equation}
where $\mu'_{i}=\mu_{obs}(z_{i})-5\ln (1+z_{i})\int_{0}^{z_{i}}\frac{1}{E(z')}dz'$.
\subsection{Baryon acoustic oscillations}
BAO is the fluctuations of the visible baryonic matter density on the length scale after the pre-recombination universe, and the BAO peak is centered on a comoving distance equal to the sound horizon at the drag epoch, $r_{s}$. BAO can be measured in the transverse and radial direction. The transverse measurement $\frac{D_{L}H_{0}}{(1+z)r_{s}}$ is sensitive to the photometric redshift, where $D_{L}$ is the Hubble-free luminosity distance shown in Eq.(\ref{25}); While the radial measurement $D_{H}/r_{s}$ is correlative to the Hubble parameter $H(z)$, where $D_{H}=\frac{c}{H_{0}E(z)}$ is the Hubble distance. The geometrical mean of radial and transverse distance named the volume averaged comoving angular diameter distance $D_{v}(z)$ is given by
\begin{equation}
D_{v}(z)=\frac{c}{H_{0}}\left[\frac{D_{L}^{2}z}{(1+z)^{2}E(z)} \right] ^{1/3},
\label{28}
\end{equation}
Then we get the observables $d(z)$ and $A(z)$ which can be written as
\begin{equation}
d(z)=\frac{r_{s}}{D_{v}(z)},
\label{29}
\end{equation}
\begin{equation}
A(z)=\frac{D_{v}(z)\sqrt{0.3 H_{0}^{2}}}{cz}.
\label{32}
\end{equation}
In this section, $H_{0}$ and $r_{s}$ are the extra parameters so we use the data of Plank15 for $H_{0}=67.3kms^{-1}Mpc^{-1}$ and $r_{s}=147.33Mpc$. And the BAO data used in this paper are listed in Table \ref{tab1}.
\begin{table}[ht]
\begin{tabular}{ccccc}
\hline
Data	 &$z$	&$d(z)$	&$A(z)$	&$D_{H}/r_{s}$\\
\hline
6dFGS\cite{beutler20116df}	&$0.106$	&$0.336\pm 0.015$	&	&\\
BOSS DR9\cite{anderson2012clustering}	&$0.57$	&$0.0732\pm 0.0012$	&	&\\
SDSS DR7\cite{percival2010baryon}	&$0.2$	&$0.1905\pm 0.0061$	&	&\\
SDSS DR7\cite{percival2010baryon}	&$0.35$	&$0.1097\pm 0.0036$	&	&\\
WiggleZ\cite{blake2011wigglez}	&$0.44$	&	&$0.474\pm 0.034$	&\\
WiggleZ\cite{blake2011wigglez}	&$0.6$	&	&$0.442\pm 0.020$	&\\
WiggleZ\cite{blake2011wigglez}	&$0.73$	&	&$0.424\pm 0.021$	&\\
BOSS DR11\cite{delubac2015baryon}	&$2.34$	&	&	&$9.18\pm0.28$\\
BOSS DR11\cite{font2014quasar}	&$2.36$	&	&	&$9.0\pm0.3$\\
SDSS DR12\cite{bautista2017measurement}	&$2.33$	&	&	&$9.07\pm0.31$\\
SDSS DR12\cite{des2017baryon}	&$2.4$	&	&	&$8.94\pm0.22$\\
\hline
\end{tabular}
\caption{The BAO data at the 1$\sigma$ level used in this paper.}
\label{tab1}
\end{table}
Next, by using the datasets \cite{beutler20116df}\cite{anderson2012clustering}, \cite{percival2010baryon}, \cite{blake2011wigglez} and  \cite{delubac2015baryon}\cite{font2014quasar}\cite{bautista2017measurement}\cite{des2017baryon}, we need to calculate the chi-squares, respectively which are written as
\begin{equation}
\chi_{1}^{2}=\sum_{i=1}^{2}\left[\frac{d_{obs}(z_{i})-d(z_{i}) }{\sigma_{i}^{2}}\right] ^{2},
\label{30}
\end{equation}
\begin{equation}
\chi_{2}^{2}=\sum_{i,j=1}^{2}\left[d_{obs}(z_{i})-d(z_{i}) \right]C^{-1}_{ij}\left[d_{obs}(z_{j})-d(z_{j}) \right],
\label{31}
\end{equation}
\begin{equation}
\chi_{3}^{2}=\sum_{i,j=1}^{3}\left[A_{obs}(z_{i})-A(z_{i}) \right]C^{-1}_{Aij}\left[A_{obs}(z_{j})-A(z_{j}) \right],
\label{33}
\end{equation}
\begin{equation}
\chi_{4}^{2}=\sum_{i=1}^{4}\left[\frac{D_{Hobs}(z_{i})/r_s-D_{H}(z_{i})/r_s }{\sigma_{i}^{2}}\right] ^{2},
\label{35}
\end{equation}
where $C^{-1}=\left( \begin{matrix}
     30124 & -17227 \\
     -17227 & 30124
\end{matrix}\right) $ and $C^{-1}_{A}=\left( \begin{matrix}
     1040.3	&-807.5 & 336.8 \\
     -807.5 & 3720.3&-1551.9\\
     336.8&-1551.9&2914.9
\end{matrix}\right) $. \\
And then we get
\begin{equation}
\chi_{BAO}^{2}=\chi_{1}^{2}+\chi_{2}^{2}+\chi_{3}^{2}+\chi_{4}^{2}.
\label{36}
\end{equation}
\subsection{Observational Hubble parameter data}
The observational methods for $H_{0}$ are the differential age method, the radial BAO size method and the gravitational wave method. In this paper, we use a compilation of 33 uncorrelated data points measured by the differential age method listed in Table \ref{tab2}.
\begin{table}[ht]
\begin{tabular}{cccc}
\hline
$z$	 &$H/km s^{-1} Mpc^{-1}$	&$\sigma/km s^{-1} Mpc^{-1}$	&Ref.\\
\hline
$0.07$&$69$&$19.68$&\cite{zhang2014four}\\
$0.09$&$69$&$12$&\cite{jimenez2003constraints}\\
$0.1$&$69$&$12$&\cite{stern2010d}\\
$0.12$&$68.6$&$26.2$&\cite{zhang2014four}\\
$0.17$&$83$&$8$&\cite{simon2005constraints}\\
$0.1791$&$75$&$4$&\cite{moresco2012improved}\\
$0.1993$&$75$&$5$&\cite{moresco2012improved}\\
$0.2$&$72.9$&$29.6$&\cite{zhang2014four}\\
$0.27$&$77$&$14$&\cite{simon2005constraints}\\
$0.28$&$88.8$&$36.6$&\cite{zhang2014four}\\
$0.3519$&$83$&$14$&\cite{moresco2012improved}\\
$0.36$&$81.2$&$5.9$&\cite{wang2017clustering}\\
$0.3802$&$83$&$13.5$&\cite{moresco20166}\\
$0.4$&$95$&$17$&\cite{simon2005constraints}\\
$0.4004$&$77$&$10.2$&\cite{moresco20166}\\
$0.4247$&$87.1$&$11.2$&\cite{moresco20166}\\
$0.4497$&$92.8$&$12.9$&\cite{moresco20166}\\
$0.47$&$89$&$50$&\cite{wang2017clustering}\\
$0.4783$&$80.9$&$9$&\cite{moresco20166}\\
$0.48$&$97$&$62$&\cite{stern2010d}\\
$0.5929$&$104$&$13$&\cite{moresco2012improved}\\
$0.6797$&$92$&$8$&\cite{moresco2012improved}\\
$0.7812$&$105$&$12$&\cite{moresco2012improved}\\
$0.8754$&$125$&$17$&\cite{moresco2012improved}\\
$0.88$&$90$&$40$&\cite{stern2010d}\\
$0.9$&$117$&$23$&\cite{simon2005constraints}\\
$1.037$&$154$&$20$&\cite{moresco2012improved}\\
$1.3$&$168$&$17$&\cite{simon2005constraints}\\
$1.363$&$160$&$33.6$&\cite{moresco2015raising}\\
$1.43$&$177$&$18$&\cite{simon2005constraints}\\
$1.53$&$140$&$14$&\cite{simon2005constraints}\\
$1.75$&$202$&$40$&\cite{simon2005constraints}\\
$1.965$&$186.5$&$50.4$&\cite{moresco2015raising}\\
\hline
\end{tabular}
\caption{The observational Hubble parameter data measured by the differential age method used in this paper.}
\label{tab2}
\end{table}
Then we need to figure out
\begin{equation}
\chi_{OHD}^{2}=\sum_{i=1}^{33}\left[\frac{H_{obs}(z_{i})-H(z_{i}) }{\sigma_{i}^{2}}\right] ^{2}.
\label{37}
\end{equation}
Finally, the total $\chi_{tot}^{2}$ is given by
\begin{equation}
\chi_{tot}^{2}=\widetilde{\chi}_{SN}^{2}+\chi_{BAO}^{2}+\chi_{OHD}^{2}.
\label{38}
\end{equation}

\begin{figure}
\centering
\includegraphics[scale=0.5]{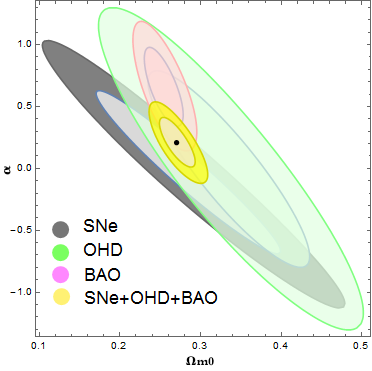}
\caption{The $1\sigma$ and $2\sigma$ level ranges of the model parameter pair $(\Omega_{m0},\alpha)$ for using SNe Ia data (grey), OHD data (green), BAO data (pink) and the combined data of SNe Ia+OHD+BAO (yellow).}\label{f3}
\end{figure}

\begin{figure}
\centering
\includegraphics[scale=0.5]{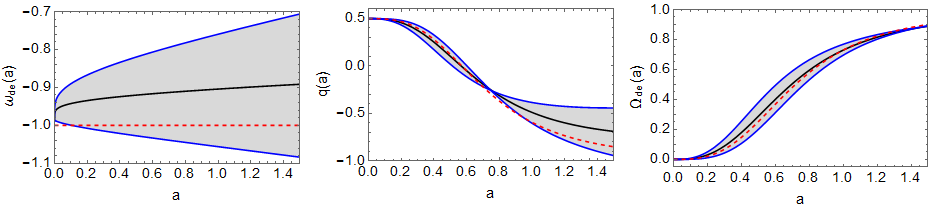}
\caption{The relations of $(a,\omega_{de})$ (left panel), $(a,q)$ (middle panel) and $(a,\Omega_{de})$ (right panel) compared with our model and the $\Lambda$CDM model. The black line and red dashed  line correspond to our model with the best-fit values listed in Table \ref{tab3} and the $\Lambda$CDM model with $\Omega_{m0}=0.27$, respectively. The shaded region and blue lines represent the 1$\sigma$ level regions and corresponding boundaries.}\label{f4}
\end{figure}

The observational constraints on the model parameter pair $(\Omega_{m0},\alpha)$ are shown in Fig.\ref{f3}; The best-fit values at the 1$\sigma$ level of parameters $\Omega_{m0}$ and $\alpha$ from
the joint constraints SNe Ia+BAO+OHD are listed in Table \ref{tab3}; The relations of $(a,\omega_{de})$, $(a,q)$ and $(a,\Omega_{de})$ compared with our model for the best-fit values which is the quintessence case and the $\Lambda$CDM model for $\Omega_{m0}=0.27$ are shown in Fig. \ref{f4}, where $q(a)$ is the deceleration parameter written as
\begin{equation}
q(a)\equiv -\frac{a\ddot{a}}{\dot{a}^{2}}=-\frac{a}{E(a)}\frac{dE(a)}{da}-1.
\label{39}
\end{equation}

From Fig.\ref{f3} and Table \ref{tab3}, we can see that the range of $\Omega_{m0}$ is acceptable and the range of $\alpha$ supports quintessence behavior slightly. But it can't be completely excluded phantom case at the 1$\sigma$ level. From Fig.\ref{f4}, the evolutional trajectories of $\omega_{de}$, $q$ and $\Omega_{de}$ can't be distinguished from the $\Lambda$CDM model at the 1$\sigma$ level. Therefore, we will adopt the $Om$ diagnostic and statefinder to discriminate our model from the $\Lambda$CDM model better. From the middle panel of Fig.\ref{f4}, we can find that the universe of our model is accelerating expansion which fits the observation. Interestingly, from the right panel of Fig.\ref{f4}, it seems that $\Omega_{de}$ of our model will gradually coincide with the $\Lambda$CDM one which tends to be a de-sitter universe. But in this model, for the quintessence case (the shaded region above the red dashed line in the right panel of Fig.\ref{f4}), if we extend $a$, we can find $\Omega_{de}$ starts to go down which is very different from the $\Lambda$CDM model. For the phantom case (the shaded region below the red dashed line in the right panel of Fig.\ref{f4}), although $\Omega_{de}$ of our model rises monotonously as same as the $\Lambda$CDM model, it will go to a little rip in the final while the $\Lambda$CDM model will go to the pseudo-rip. The detail of rip will be discussed at the rip section below.
\begin{table}[ht]
\begin{tabular}{c|cc}
\hline
&SNe Ia+OHD+BAO&\\
\hline
$\Omega_{m0}$&$0.270^{+0.039}_{-0.034}$\\
$\alpha$&$0.210^{+0.328}_{-0.332}$\\
\hline
\end{tabular}
\caption{The best-fit values at the 1$\sigma$ level of parameters $\Omega_{m0}$ and $\alpha$ from
the joint constraints SNe Ia+BAO+OHD.}
\label{tab3}
\end{table}
\section{Discrimainations BY \texorpdfstring{$Om(a)$}{} diagnostic and the statefinder}\label{sec6}
As more and more dark energy models are proposed so far, how to discriminate different dark energy models becomes an important and meaningful issue. In the first part of this section, we employ $Om(a)$ diagnostic to distinguish our model with the best-fit values from the $\Lambda$CDM model. In the second part, we use the statefinder parameters to discriminate among the quintessence picture, the phantom picture and the $\Lambda$CDM model.
\subsection{\texorpdfstring{$Om(a)$}{} Diagnostic}
The $Om(a)$ diagnostic\cite{sahni2008two} is a geometrical method which combines Hubble parameter and redshift to discriminate the dark energy models by measuring their deviation from the $\Lambda$CDM model. $Om$ is defined as
\begin{equation}
Om(a)=\frac{E^{2}(a)-1}{a^{-3}-1}.
\label{40}
\end{equation}
For a spatially flat $\Lambda$CDM model, $E^{2}(a)=\Omega_{m0}a^{-3}+(1-\Omega_{m0})$. So $Om(a)|_{\Lambda CDM}-\Omega_{m0}=0$ which provides a null test of $\Lambda$CDM hypothesis.\\
\begin{figure}
\centering
\includegraphics[scale=0.5]{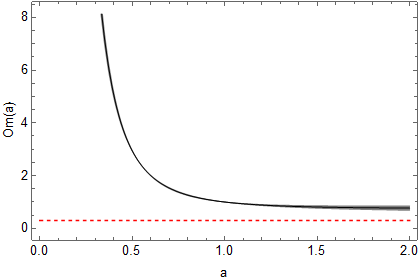}
\caption{The $Om$ diagnostic for our model and the $\Lambda$CDM model. The black line represents our model with the best-fit values listed in Table \ref{tab3}.  The red dashed line represents the $\Lambda$CDM  model with $\Omega_{m0}=0.3$. The shaded region represents the 1$\sigma$ level regions.}\label{f5}
\end{figure}
In Fig.\ref{f5}, we plot the evolutional trajectories of our model and the $\Lambda$CDM model. From Fig.\ref{f5} we can see that our model has a relatively large deviation from the $\Lambda$CDM model at high redshifts and gradually approaches the $\Lambda$CDM model at low redshifts and in the future evolution. But they can be easily distinguished from each other at the 1$\sigma$ level all along. The $Om$ diagnostic discriminates our model from the $\Lambda$CDM model very well.
\subsection{Statefinder}
The $Om(a)$ diagnostic relies on the first order derivative of the scale factor with the respect to cosmic time alone while the statefinder\cite{sahni2003statefinder} relies on the  higher order derivatives. The geometric parameter pair $(r,s)$ are defined as
\begin{equation}
r\equiv\frac{\dddot{a}}{aH^{3}},
\label{41}
\end{equation}
\begin{equation}
s\equiv\frac{r-1}{3(q-\frac{1}{2})},
\label{42}
\end{equation}
where $q$ is the deceleration parameter shown in Eq.(\ref{39}). By using Eqs.(\ref{41}) and (\ref{42}) we can derive $q$ and $s$ of our model and the $\Lambda$CDM model, and they are listed in Table \ref{tab4}. For the better comparison, we also list the dimensionless Hubble parameter $E$, the density ratio parameter of the matter $\Omega_{m}$ and the deceleration parameter $q$ in Table \ref{tab4}. 
\begin{table}[ht]
\begin{tabular}{c|cc}
\hline
&our model&the $\Lambda$CDM model \\
\hline
$E^{2}$&$1-\Omega_{m0}-\alpha \ln a+ \Omega_{m0}a^{-3}$&$1-\Omega_{m0} + \Omega_{m0}a^{-3}$\\
$\Omega_{m}$&$\frac{\Omega_{m0}a^{-3}}{1-\Omega_{m0}-\alpha \ln a+\Omega_{m0}a^{-3}}$&$\frac{\Omega_{m0}a^{-3}}{1-\Omega_{m0}+\Omega_{m0}a^{-3}}$\\
$q$&$-1+\frac{3}{2}\Omega_{m}+\frac{\alpha}{2E^{2}}$&$-1+\frac{3}{2}\Omega_{m}$\\
$r$&$1-\frac{3\alpha}{2E^{2}}$&$1$\\
$s$&$\frac{\alpha}{3E^{2}-3\Omega_{m0}a^{-3}-\alpha}$&$0$\\
\hline
\end{tabular}
\caption{The comparison of different parameters between our model and the $\Lambda$CDM model.}
\label{tab4}
\end{table}
\begin{figure}
\centering
\includegraphics[scale=0.6]{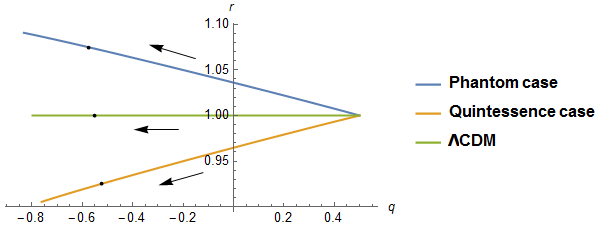}
\caption{The statefinder pair $(q,r)$ for quintessence case (blue line), phantom case (orange line) and the $\Lambda$CDM model (green line). Arrows represent the directions of time evolution. The spots indicate the present epoch. We have used $\Omega_{m0}=0.3$, $\alpha=0.05$ for quintessence case, $\Omega_{m0}=0.3$, $\alpha=-0.05$ for phantom case and $\Omega_{m0}=0.3$, $\alpha=0$ for the $\Lambda$CDM model.}\label{f6}
\end{figure}
\begin{figure}
\centering
\includegraphics[scale=0.5]{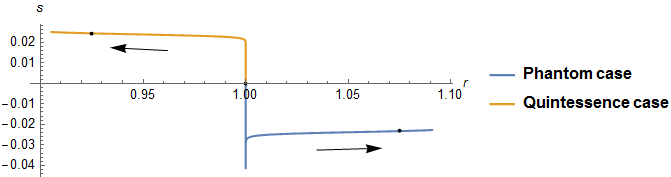}
\caption{The statefinder pair $(r,s)$ for quintessence case (blue line), phantom case (orange line) and the $\Lambda$CDM model (the fixed point $(1,0)$). Arrows represent the directions of time evolution. The spots indicate the present epoch. We have used $\Omega_{m0}=0.3$, $\alpha=0.05$ for quintessence case and $\Omega_{m0}=0.3$, $\alpha=-0.05$ for phantom case.}\label{f7}
\end{figure}
Fig.\ref{f6} shows the relation between $q$ and $r$. The relation between $r$ and $s$ is shown in Fig.\ref{f7}. Both of two figures indicate that the quintessence case and the phantom case can be well distinguished from the $\Lambda$CDM model and will gradually deviate from each other. Interestingly, in Fig.\ref{f7}, when two cases deviate slightly from $a=0$, they both oscillate up and down at point (1,0) and constantly overlap. Then they quickly move away from point (1,0) in the opposite directions and immediately tend to be stabilized and part ways. It implies that this two cases may share the same phase at the birth of the universe.
\section{The rip}\label{sec7}
From the conservation equation $\dot{\rho}=-3H\rho(1+\omega)$ we know that the density will increase in the future when the EoS of dark energy $\omega_{de}<-1$ which corresponds the phantom case. Based on various evolutionary behaviors of $H(t)$, we divide the ultimate fates of the universe into the following categories\cite{frampton2012pseudo}: (1) The big rip, for which $H(t)\rightarrow\infty$ at finite time. At that time, the dark energy density is infinity and produces an infinite repulsion, the gravitationally bound system will be dissociated  in order of large to small\cite{caldwell2003phantom}. (2) The little rip, for which $H(t)\rightarrow\infty$ at infinite time. This scenario has no singularity in the future whereas also leads to a dissolution of bounds tructures at some point in the future\cite{frampton2011little}. (3) The pseudo-rip, for which  $H(t)\rightarrow constant$ which is an intermediate case between the de-Sitter cosmology and the little rip. Next, we will make a rip analysis for the phantom case of our model briefly. \\
For our model, the Hubble parameter is
\begin{equation}
H^{2}=(\frac{\dot{a}}{a})^{2}=\frac{8\pi G}{3}\rho_{0}(1-\Omega_{m0}-\alpha\ln a+\Omega_{m0}a^{-3}),
\label{43}
\end{equation}
and $\alpha<0$ for the phantom case. When $a\rightarrow\infty$, Eq.(\ref{43}) can be simplified as
\begin{equation}
(\frac{\dot{a}}{a})^{2}\rightarrow\frac{8\pi G}{3}\rho_{0}(-\alpha\ln a)\equiv n\ln a,
\label{44}
\end{equation}
where $n\equiv-\frac{8\pi G}{3}\rho_{0}\alpha$. By solving the differential Eq.(\ref{44}), we can obtain the scale factor $a$ as a function of time $t$
\begin{equation}
a=\exp [n(t-t_{0})^{2}/4],
\label{45}
\end{equation}
where $t_{0}$ is the present value of time. Substitute Eq.(\ref{45}) to Eq.(\ref{44}), we get
\begin{equation}
H=\frac{1}{2}n(t-t_{0}).
\label{46}
\end{equation}
From Eq.(\ref{46}) we can find $H(t)\rightarrow\infty$ as time goes to infinity. So the ultimate fate of the phantom case of our model is the little rip. 
\section{Conclusions and discussions}\label{sec8}
In this paper, we propose a pressure parametric model of the total energy components in a spatially flat FRW universe. This model has two parameters $\Omega_{m0}$ and $\alpha$ where $\Omega_{m0}$ is the present-day dark matter density parameter and $\alpha$ displays the model difference from the flat $\Lambda$CDM model.  By constraining with the datasets of SNe Ia, BAO and OHD, we find that $\Omega_{m0}=0.270^{+0.039}_{-0.034}$ and $\alpha=0.210^{+0.328}_{-0.332}$ at the 1$\sigma$ level which means our universe slightly biases towards quintessence behavior while it can not be completely excluded phantom at the 1$\sigma$ level. And it also implies that our model includes the $\Lambda$CDM model when $\alpha=0$. Then we use $Om(a)$ diagnostic to discriminate our model with the best-fit values from the $\Lambda$CDM model. We find that our model deviates relatively far from the $\Lambda$CDM model at high redshifts and gradually approaches the $\Lambda$CDM model in the future. However they can be easily distinguished from each other at the 1$\sigma$ level all along. Next, we use the statefinder to discriminate among the quintessence case, the phantom case and the $\Lambda$CDM model. Both of panels $(q,r)$ and $(r,s)$ indicate that quintessence and phantom scenarios can be well distinguished from the $\Lambda$CDM model and will gradually deviate from each other. Finally, we discuss the fate of universe evolution named the rip analysis for the phantom case of our model and find that the universe will run into a little rip stage which has no singularity in the future whereas also leads to a dissolution of bound structures at some point in the future.

On the one hand, dark energy phenomenon has appeared about two decades, but we still do not know its physical reality. While waiting for upcoming new observations, lots of theoretical efforts need continuously paid with the hope we can understand it better. On the other hand, the constraints give a tiny $\alpha$, so this model can also provide a possible solution for other studies to approximate the pressure at low redshifts.

\section*{Acknowledgments}
Jun-Chao Wang thanks Yan-Hong Yao and Yang-Jie Yan for the fruitful discussions. This study is supported in part by National Natural Science Foundation of China.

\bibliographystyle{spphys}
\bibliography{bibfile}
\end{document}